# Porous metal nitride film synthesis without template

Adrien Baut[1], Michael Pereira Martins[1] and Andreas T. Güntner[1*]

[1] Human-centered Sensing Laboratory, Department of Mechanical and Process Engineering, ETH Zurich, CH-8092 Zurich, Switzerland.

[*] Corresponding author: andregue@ethz.ch

# Abstract

Metal nitrides possess exceptional catalytic, electronic and physical properties making them widely used in (opto-)electronics and as hard coatings. When used as films in surface-active applications, however, their performance remains limited by poor mass transfer and reduced accessibility of reactive sites. This is associated to compact film architecture yielded by conventional deposition techniques (e.g., 16-26 % for sputtered $W_2N$). Here, we demonstrate a template-free method for the design of highly porous (porosity > 84 %) metal nitride films with high compositional versatility, as demonstrated for $Cu_3N$, $W_2N$, $MoN_x$ and TiN. These are obtained by exploiting the self-assembly of fractal-like metal oxide agglomerates during deposition from aerosols followed by their dry nitridation. In case of $Cu_3N$, monocristalline oxide nanoparticles were converted to polycrystalline nitrides during nitridation, as traced by X-ray diffraction and electron microscopy. Such films feature consistently lower resistances than their metal oxide counterparts, as well as high reactivity and mass transfer. This is exploited examplarily for molecular sensing of $NO_2$ at only 75 °C temperature, leading to up to a five-fold higher response with faster response time over more compact spin-coated films for. As a result, our approach overcomes critical mass transfer performance limitations of metal nitride films that are also relevant for other applications like electrocatalysis and energy storage.

# Keywords

# Introduction

Metal nitrides (MN) are distinguished by the coexistence of covalent and ionic bonding within their lattice[1,2]. In transition MN, this bonding leads to the contraction of the *d*-band, which increases the density of electrons near the Fermi level[2,3]. As a result, MN feature beneficial bulk properties such as excellent electrical conductivity and high thermal stability, making them highly valuable for use in protective coatings[4], electronics[5], and LED manufacture[6], to name a few examples. Additionally, the presence of both acidic and basic sites on their surface[7] yields high catalytic activities[8], which makes them attractive for electrochemical energy conversion[9], energy storage[3], ammonia production[10], and molecular sensing (e.g., of air pollutants including $NO_2$[11]). As catalytic activity increases with accessible surface area, small particles as well as porous film architectures are desirable[12].

MN nanoparticles are usually synthesized using wet chemistry (e.g., $Cu_3N$[13]) or by exposing solid precursors (e.g., metal or metal oxide) to a source of nitrogen ($NH_3$[14,15], $N_2$[16], $Mg_3N_2$[17], or urea[18]) at high temperatures. The resulting MN crystal sizes obtained with the latter route typically range between 5-100 nm depending on the initial precursor material[13,14,16-19]. For instance, initial $TiO_2$ particle size between 17-98 nm yielded 12-17 nm large TiN particles[17]. MN films have been obtained by spin coating[20], chemical vapor deposition[21], sputtering[22], or atomic layer deposition[23]. Yet, these techniques are known from other nano-structured materials to result in rather dense films, for example, porosities of 18-54 %[24,25] when spin-coating sol layers or even lower for sputtered $W_2N$ (porosities of 16[22] and 26 %[26]), $Cu_3N$ (calculated porosity of 13 %[27]) and TiN coating films (< 1 %[28]). These compact film architectures are ineffective for applications relying on high mass transfer and surface reactions, as demonstrated with lower catalytic conversion[29] or slower sensor responses for molecular sensing[30].

Higher porosities were achieved, for instance, by using nanocasted[31,32] or self-assembled[33,34] templates. As an example, a series of $SiO_2$ and $TiO_2$ mesoporous films were structured by adding different concentrations of block copolymers Pluronic F127 or P123, achieving up to 68 %[35] porosities. Other examples of templated porous MN include zeolites[36], mesoporous silica (e.g., SBA-15[31]) or diatomite[37] (porosity of 44-66 %[38]). However, such templates require close control over process conditions (e.g., humidity and temperature[34]) to prevent the collapse of the structure[34] and the template may need to be removed with hazardous compounds (e.g., with HF[37]). Such templating usually takes few hours to several days[33,39] and can result in cracks, even when optimal parameters are used[33].

In this work, we investigate a template-free pathway to yield highly porous and homogeneous MN films (Figure 1). The films are obtained via direct deposition of metal oxide (MOx) agglomerates from aerosols, followed by dry conversion to MN with gaseous $NH_3$. The versatile concept is demonstrated for $Cu_3N$, TiN, $W_2N$, and $MoN_x$ films that are characterized by electron microscopy to assess the porosity. X-ray diffraction and microscopy imaging is applied to monitor the crystal phase dynamics and to reveal the conversion mechanism. The electronic properties and molecular sensing capabilities of such highly porous MN films are assessed to demonstrate immediate practical impact for selective trace-level detection of air pollutants, as demonstrated exemplarily with $NO_2$.

## Results and Discussion

*Template-free synthesis of highly porous metal nitride films*

Porous MN films were formed by first dispersing and combusting liquid precursor solutions containing the target metal ions mixed on a molecular level[40,41] (Figure 1a). While we focus here on monometallic MN, also the mixing of two (e.g., Co/Cu[42]) and three (Sn/Sb/Pd)[43] metals has been demonstrated for oxides and noble metals. Particles are obtained by nucleation, coagulation, and sintering[44] (Figure 1a) to form nanocrystals that can be even outside their thermodynamically stable configuration (e.g., ε-$WO_3$[45], $In_4Sn_3O_{12}$[46] or $CuCo_2O_3$[47]). The resulting nanoparticles agglomerate to fractal-like building blocks that self-assemble to highly porous films when deposited onto water-cooled substrates by thermophoresis[48] (Figure 1b). As shown exemplarily on CuO by scanning electron microscopy (SEM) in Figures 2a,b, such films feature highly porous (i.e., 98 ± 1 %) architecture, in agreement with literature for CuO[30] and other MOx[48-51]. The film thickness is 8.5 ± 0.7 µm (Figure S1a) that can easily be controlled through deposition time[48,49]. The agglomerated state of the particles can be seen in Figure 2c.

These MOx films were converted to MN by gaseous $NH_3$ (Figure 1c). Due to the high porosity of the MOx film, the $NH_3$ can access even the deepest layers of the film and convert these uniformly while hardly affecting the film architecture (Figure 2d). In fact, the resulting $Cu_3N$ films feature a porosity of 91 ± 1 % (Figures 2d,e) and a thickness of 5.7 ± 0.4 µm (Figure S1b). The constituent particles of the film conserve their agglomerated state, as observed in Figure 2e,f. The same process has been applied to TiN, $W_2N$, and $MoN_x$ to show the versatility of our template-free dry synthesis for porous MN (cross-sectional SEM images in Figures S2a-c). Consistently high (≥ 84 %, filled black symbols, Figure 2g) porosity was obtained, quite similar to the corresponding MOx films (empty symbols). Such porosities are by far superior to previous $Cu_3N$[27], TiN[28], and $W_2N$[22,26] films made by sputtering (≤ 26 %, right-half full symbols in Figure 2g) and even outperform significantly more tedious templated methods (≤ 68 % for $SiO_2$ and $TiO_2$[35]). For a direct comparison, we prepared $Cu_3N$ films also via traditional spin-coating (see Methods) from our combustion aerosol-made and filter-collected CuO powders (Figure 1d) and converted them to $Cu_3N$ (Figure 1e, Figure S3). Also, this film features a significantly smaller (51 ± 9 %) porosity than the directly deposited ones.

To elucidate the crystal conversion efficiency of our process, we show the X-ray diffraction (XRD) patterns of CuO, TiO$_2$, WO$_3$ and MoO$_3$ powders and the corresponding MN after dry conversion in Figure 2h. In general, the as-prepared MOx particles are highly crystalline as evidenced by the distinct XRD peaks (empty symbols) with sometimes multiple phases, in agreement with literature[49,50,52,53]. Importantly, the MOx nanocrystals are completely converted to pure Cu$_3$N (circles), TiN (diamonds), and W$_2$N (triangles), as no MOx-related peaks are visible anymore. In the case of molybdenum, even two phases of hexagonal MoN (inverted triangle) and cubic Mo$_2$N (star) are obtained.

*Nanocrystals and conversion dynamics*

To gain more insights into the conversion mechanism and crystal growth of the nanoparticles, we further explored the CuO-Cu$_3$N system. When exposing flame-made powders as packed bed to NH$_3$ at 200 °C for 0.25 h to 5 h, the conversion of the oxide nanoparticles was followed via XRD in Figure 3a. Already after 0.25 h, the main peak of Cu$_3$N at $2\theta = 23.5°$ (filled circles) emerges while those of CuO (empty circles) at 35.5° and 38.7° are significantly decreased. The calculated Cu$_3$N content was 65 wt% (Figure 3b, squares, right ordinate), as determined by Rietveld refinement. As dwell time increases, more CuO is converted until the related XRD peaks disappear completely after 5 h, indicating full conversion to Cu$_3$N. Importantly, no additional phases were observed, in contrast to literature on wet synthesis of Cu$_3$N, where additional metallic Cu and Cu$_2$O were found for prolonged conversion times[54,55]. Their presence is attributed to the decomposition of Cu$_3$N into Cu and its further reoxidation into Cu$_2$O due to the surrounding medium[54,55].

The reduction of the CuO peak intensities in Figure 3a is in agreement with a shrinking crystal size of CuO (Figure 3b, open circles) from initially $6.4 \pm 0.3$ nm until it is not detected anymore after full conversion. In contrast, the Cu$_3$N crystal size (Figure 3b, red circles) grows to final $3.9 \pm 0.2$ nm after 5 h. Note that our process is rather reproducible, as three identically prepared powders of CuO and Cu$_3$N yielded crystal size variations below 5 % for both phases (error bars in Figure 3b before and after full conversion, hidden behind symbols).

For the initial CuO, the crystal and particle sizes ($d_{XRD}$ = 6.4 nm and $d_{BET}$ = 8.3 nm) are rather similar, indicating monocrystallinity, in line with literature[53]. This observation is supported by high resolution transmission electron microscopy (HRTEM) images of those particles (Figure 3c), where visible lattice fringes of a crystal occupy entire particles. The corresponding fringes feature a spacing of 0.23 nm, attributed to the (111) orientation (see inset of Figure 3c), in line with selected area electron diffraction (SAED) pattern in Figure S4a. Most interestingly, the final (i.e., after 5 h dwell time, Figure 3b) crystal size is clearly smaller than the particle size ($d_{XRD}$ = 3.9 nm vs. $d_{BET}$ = 12.7 nm), suggesting polycrystallinity. In fact, multiple crystal domains within a single particle were observed by HRTEM (Figure 3d) with lattice spacing of 0.22 (blue) and 0.38 nm (red), similar to the (111) and (100) orientations of Cu$_3$N. An SAED pattern in Figure S4b confirms these crystal directions, along with the presence of the (110) orientation. The polycrystallinity of Cu$_3$N particles has also been observed via wet synthesis[13] or dry nitridation of Cu containing salts[56].

As crystal and particle sizes significantly affect performance (e.g., sensitivity in chemoresistive sensing[57]), controlling it during synthesis is critical. Figure 3e shows the Cu$_3$N crystal size when varying the temperature during a 2 h nitridation with the corresponding XRD patterns provided in Figure S5. The crystal size increases quite linearly when just altering temperature between 200-310 °C, allowing for good control between ca. 4-7 nm. This is associated with sintering and aggregation effects, as observed for TiN particles before[15]. Note that even a larger range of Cu$_3$N size should be accessible by controlling initial CuO crystal size, that can be above 40 nm by enclosing the spray flame during synthesis with a quartz tube[53].

*Electronic and low temperature molecular sensing properties of porous MN films*

To investigate the electronic properties of our porous films, we first measured the bandgaps of each MOx-MN pair using the Tauc plot method with the Kubelka-Munk function $F(R_\infty)$. The data obtained is exemplarily shown in Figure 4a for $TiO_2$ (black, left ordinate axis) and TiN (red, right ordinate axis), where the obtained direct bandgaps are 3.5 and 1.8 eV respectively, in accordance with literature (3.47 eV for the direct transition of P25-$TiO_2$[58] and 1.58-1.95 eV for TiN[59]). The Tauc plots of the other MOx-MN pairs can be found in Figure S6. The MN feature consistently lower bandgaps than MOx (Figure 4b), except $Cu_3N$ that was quite similar to CuO. This has been associated to the lower electronegativity of nitrogen compared to oxygen[60]. As a result, electrons are less localized in MN than in MOx, which leads to lower bandgaps for MN[61] as observed on a variety of materials (e.g., binary[62,63], ternary[62] semi-conductor and metal oxides semiconductors specifically[64]. The modification of the lattice structure from the MOx to MN, as well as the interatomic distance between the metal atoms and the oxygen/nitrogen atoms in the crystal may affect the bandgap as well[65].

Lowering the bandgap has shown improvement in catalytic performances by reducing the activation energy of certain chemical reactions[66] or enabled photocatalysis in visible light[67], rendering MN more attractive for these applications. Additionally, charge carriers are more easily thermally excited in lower bandgap materials, leading to sometimes orders of magnitude lower resistances for similar architectures. This observation is in agreement with our findings that MN films (red symbols in Figure 4b) feature consistently lower resistances than their MOx counterparts (black symbols), when both fabricated by direct deposition (Figure 1).

Next, we assessed the chemical and electronical properties of porous MN films for molecular $NO_2$ sensing at 75 °C. $NO_2$ is a major environment pollutant with strict exposure limits (e.g., annual exposure of 20 parts-per-billion (ppb) in the E.U.[68] that are expected to halve by 2030[69]). Figure 4c shows the resistance of the directly deposited porous $Cu_3N$ sensor (red, left ordinate) during an exposure of 1000 ppb $NO_2$ under realistic 50 % relative humidity (RH) in air. The sensor features a baseline resistance around 3.5 kΩ that is lower than other $NO_2$ sensors operated even at higher temperatures (e.g., 70 MΩ for a similarly directly deposited $WO_3$ sensor at 125 °C[49]). When exposed to $NO_2$, a p-type behaviour is observed as resistance decreases to 0.7 kΩ when exposed to this oxidizing analyte, in accordance with the literature[70]. The corresponding sensor response is 2.95 with response and recovery times of 11 and 99 min, respectively.

In comparison, a denser spin-coated sensor (porosity of 51 % vs. 91 % for directly deposited) shows a lower baseline resistance of 57 Ω (Figure 4c, blue, right ordinate) that is attributed to less grain contact resistances due to more compact structures[71] (compare Figures 2e and S3b). Most importantly, the sensing performances of the spin-coated film features four times smaller response (0.73 vs. 2.95) to 1000 ppb $NO_2$ and four times slower response time. As both films feature comparable thicknesses (5.7 ± 0.4 µm from for the spin-coated film, Figure S1b, and 5.0 ± 1.7 µm for the directly deposited film, Figure S3a) and $Cu_3N$ particle sizes (Figure 2f vs. Figure S3b), the superior performance of the directly deposited film is attributed to higher porosity. Higher porosity facilitates faster mass transfer of $NO_2$ inside the film and it provides a better accessibility of reactive sites on the $Cu_3N$ surface, as observed in sensing[30] and photocatalysis[29].

*Applicability of porous $Cu_3N$ films as $NO_2$ sensors*

To demonstrate immediate practical impact for our process, we challenged the porous $Cu_3N$ sensor further for the detection of $NO_2$ over a range of relevant concentrations between 50-1000 ppb (Figure 4d) in air at 50% RH. Remarkably, the sensor clearly resolves concentrations down to 50 ppb, meeting the E.U. 1-hour exposure limit of 100 ppb[68]. The high signal-to-noise ratio (SNR) at 50 ppb (SNR >100, raw data in Figure S7) leads to a linearly extrapolated (i.e. theoretical) limit of detection at SNR = 3 of 0.1 ppb. This is five times lower than for a $WO_3$ sensor operated at 125 °C[49] and around three orders of magnitude lower than for a $MoS_2$ sensor at room temperature[72], rendering our porous $Cu_3N$ sensor promising for checking adherence to upcoming lower exposure limits[69].

Outdoor and indoor air contain approximately 250 compounds[73], so high selectivity to $NO_2$ is a necessary requirement for a functional sensor. Figure 5b shows the response of the $Cu_3N$ sensor to a wide range of critical confounding analytes (alcohols, ketones, $H_2$, aromatic compounds, and NO) at a concentration of 1000 ppb. Remarkably, the sensor responds mainly to $NO_2$, less to NO (selectivity of 3.6), and hardly to any other compound with selectivities between 420 and more than $10^4$ for $H_2$ and benzene, respectively. These selectivities are superior to other $NO_2$ sensors like $WS_2$ (263 towards ethanol[74]) and comparable to $SnS_2$[75] ($> 10^4$ towards benzene and acetone) or $WO_3$[49] ($> 10^4$ towards acetone and $H_2$ and $> 10^5$ towards ethanol), though the latter two had to be heated to 200 °C and 125 °C, respectively. The low operating temperature coupled with a high selectivity to volatile organic compounds renders this porous $Cu_3N$ attractive for air quality monitoring devices. Such type of sensors are already compatible with handheld devices[76]. Filters[77,78] may enhance selectivity further, if needed, and their combination with other sensors in arrays[79,80] enables multi-tracer capabilities (e.g., with benzene[81] or formaldehyde[82]).

Another challenge is humidity for MOx sensors, with responses usually decreasing with increasing RH (e.g., $WO_3$ with $NO_2$ due to oxidation of oxygen vacancies[83]). Remarkably, the response of the porous $Cu_3N$ increases from 1.06 in dry air to 3.13 at 75% RH (Figure 5c). This suggests that presence of water vapor in the vicinity of the $Cu_3N$ surface promotes the surface and/or transduction, though further studies will be needed to clarify this effect. Finally, we also assessed the stability of the sensor over 9 days (Figure 5d) when exposed daily to 1000 ppb of $NO_2$ at 75 °C and 50% RH in air. The sensor response was on average 3.2 with a standard deviation of 0.7.

# Conclusion

A template-free synthesis method for MN films was introduced that yielded exceptional porosities ≥ 84 %, significantly higher than conventional methods, even when using templates. By direct deposition of MOx nanoparticles through thermophoresis and subsequent dry nitridation, porous film architecture was obtained and preserved by conversion on the particle level. Such films constitute nanoparticulate morphology, that can be refined during nitridation through formation of polycrystallinity, as shown in the case of $Cu_3N$. The approach was versatile, as demonstrated with various metals and it allows controlled film thickness and crystal size optimization. Such porous MN films feature low resistance and high mass transfer, that is attractive for molecular sensing, heterogeneous catalysis or energy storage, to name a few examples.

Immediate practical impact was demonstrated with porous $Cu_3N$ films that detected trace-level $NO_2$ concentrations at moderate temperature. Such sensors exhibited excellent selectivity over critical compounds and enhanced performances under humidity, all of which are challenges of the well-established chemoresistors. Together with good stability over time, these features indicate potential for detection of $NO_2$ in real-world air quality monitoring scenarios.

# Methods

*Metal oxide nanoparticles and films*

MOx nanoparticles of various composition were produced with an FSP reactor, described in detail elsewhere[41]. In brief, previously reported precursor solutions[45,50,53,84] were fed through the nozzle of the FSP burner at the pressure drop, precursor and $O_2$ dispersion flowrates specified in Table S1. A ring-shaped $CH_4/O_2$ flamelet (1.25/3.25 L/min) ignited and sustained the resulting precursor spray, while additional $O_2$ was introduced as sheath gas at 5

L/min through a surrounding porous sinter-metal plate. As prepared nanoparticles were collected onto a water-cooled glass microfiber filter (GF-6 Albet-Hahnemuehle, 257 mm diameter) positioned 57 cm above the nozzle with the help of a vacuum pump. The powder was scraped off the filter and sieved (250 μm mesh) to remove filter residues.

Porous films were formed by direct deposition of flame-made nanoparticles through thermophoresis[48] onto $Al_2O_3$ sensor substrates. These substrates (electrode type #103, Electronic Design Center, Case Western University, USA) featured interdigitated Pt electrodes on the front and a Pt heater on the back. The substrates were fixed onto a water-cooled holder positioned 20 cm above the nozzle. The deposition times are indicated in Table S1 as well. Note that the films were annealed in an oven (CWF 1300, Carbolite Gero, Germany) for 5 h at 500 °C to increase the adhesion to the substrate.

*Metal nitride nanoparticles and films*

The MN powders were obtained by the nitridation of a packed bed containing 60-80 mg (Table S2) as-prepared MOx nanoparticles that were kept in place with glass wool[85] (Sigma-Aldrich, Switzerland) inside a quartz tube (1cm inner diameter, 70 cm long). The quartz tube was then fixed inside a tube furnace (CTF 12/65/550, Carbolite, Germany) and heated with heating rate, target temperature and dwell time specified in Table S2 (unless otherwise stated in the text). The flow of pure $NH_3$ (PanGas, individual flow rates in Table S2) was controlled using high-precision mass flow controllers (Bronkhorst, Netherlands) and kept constant during the entire conversion. The conversions of the directly deposited MOx films were done using the same parameters as for the powders, replacing the packed beds of powder by the directly deposited annealed sensors.

For comparison, more compact films of $Cu_3N$ were prepared by spin-coating. For that, 4 mg of flame-made as-prepared CuO nanoparticles were mixed with 180 μL of 1,2-propanediol (Sigma-Aldrich, Switzerland) and ultrasonicated to fully disperse the nanoparticles with the solvent. Afterwards, the solution was pipetted onto the sensor substrates, covering the whole surface, and spun at 300 rpm for 30 minutes with an acceleration of 100 rpm/s. After spinning, the substrates with the solution were annealed in an oven (CWF 1300, Carbolite Gero, Germany) at 500 °C for 5 h to evaporate the solvent and reinforce the adhesion of the films to the substrates. Finally, the films were converted using the same protocol as for the directly deposited sensor (see Table S2).

*Powder and film characterization*

Crystallographic characterization of powders was performed by XRD using a Bruker D2 Phaser (Bruker, USA). The samples were scanned with Cu $K_\alpha$ radiation at 30 kV and 10 mA from 10 to 70° with a scanning speed of 1.1 s/step and a scanning step size of 0.012°. The crystal phases were identified using the Diffrac.eva V3 (Bruker, USA) software with the following reference crystal patterns: monoclinic CuO (PDF- 72-0629), cubic $Cu_3N$ (PDF-74-0242), tetragonal anatase (PDF-21-1272) and tetragonal rutile (PDF-21-1276) $TiO_2$, cubic TiN (PDF-38-1420), orthorhombic $MoO_3$ (PDF-05-0508), monoclinic $MoO_3$ (PDF-47-1081), monoclinic $MoO_2$ (PDF-78-1069), orthorhombic $Mo_4O_{11}$ (PDF-72-0448), cubic $Mo_2N$ (PDF-25-1366), hexagonal MoN (PDF-77-1999), monoclinic $\gamma$-$WO_3$ (PDF-83-0950), monoclinic $\varepsilon$-$WO_3$ (PDF-24-0747) and cubic $W_2N$ (PDF-25-1257). The crystal sizes and weight contents were determined via the Rietveld refinement method[86] from the TOPAS 4.2 (Bruker, USA) software.

The optical bandgap of the powders was determined from UV-Vis spectra using the Tauc plot method with the Kubelka-Munk function. For that, the MN and MOx powders were mixed with barium sulfate (Sigma-Aldrich, Switzerland) with a ratio of 1:9 by weight to obtain a total of 180 mg. The spectra were measured with a UV-Vis-NIR spectrophotometer (Cary 5000 Agilent, USA) operated for wavelengths between 200 – 3000 nm at a scanning rate of 1 nm/s.

The specific surface area (SSA) was obtained via nitrogen adsorption at 77 K (Tristar II Plus, Micromeritics, USA) with a Brunauer-Emmett-Teller (BET) 8-point method. Prior to the measurement, the powders were degassed under nitrogen atmosphere for 1 h at 100 °C. The equivalent diameters ($d_{BET}$) were determined assuming spherical particles, with the density being calculated using the nominal amount of CuO and $Cu_3N$ derived from the XRD patterns and the density of CuO (6.40 g/cm³) and $Cu_3N$ (5.84 g/cm³).

High resolution transmission electron microscopy particle images were obtained on a Grand-ARM300F (JOEL, Japan), equipped with a cold field emission gun operated at 300 kV. The aberration corrections enable atomic-resolution imaging in TEM mode. The film morphology was investigated by SEM using a Hitachi S-4800 FE-SEM. The film thickness ($s_{sl}$) was obtained with > 100 measurements per deposited film using the ImageJ software on a cleaved substrate (examples in Figure S1). The porosity of the films was calculated by attenuation of the X-ray signal of the $Al_2O_3$ peaks from the substrate due to the nanoparticle film[48]. For that, the peak intensity at 35.1, 43.3, and 57.5° of the bare ($I_{in}$) and covered ($I_{em}$) substrate were measured. The radiation pathway, $s_s^*$, was then determined using the exponential attenuation law[87]:

$$\frac{I_{em}}{I_{in}} = \exp[-\frac{\mu}{\rho_s}\rho_s s_s^*]$$

with $\mu/\rho_s$ being the mass attenuation coefficient which can be obtained from the XCOM Photon Cross Section Database[88] and $\rho_s$ the material density. The used values of density were $TiO_2$ (3.78 g/cm³), TiN (5.4 g/cm³), $WO_3$ (7.16 g/cm³), $W_2N$ (11.51 g/cm³), $MoO_3$ (4.69 g/cm³) and $MoN_x$ (9.31 g/cm³), while CuO and $Cu_3N$ have been specified above. The corrected bulk film thickness, $s_s$, was then obtained by considering the radiation pathway for the incident angle and taking half of the value since it is twice the bulk film thickness. The porosity is finally calculated using the deposited film thickness ($s_{sl}$) from SEM and the corrected bulk film thickness from XRD:

$$\text{Porosity} = 1 - \frac{V_s}{V_{sl}} = 1 - \frac{s_s}{s_{sl}}$$

*Gas Sensing Performance*

The deposited films were mounted onto Macor holder, inserted in a Teflon chamber and tested with a gas mixing setup specified elsewhere[50]. The film resistances between the substrate's interdigitated Pt electrodes were measured using a multimeter (Keithley, Integra Series 2700, USA, for resistances below 100 MΩ; Keithley, DMM7510 7, USA, for resistances between 100 MΩ and 1.2 GΩ). The resistance of the films was measured while exposed to a flow of 50 ml/min of dry air (PanGas, 5.0, $C_nH_m$ and $NO_x \leq 100$ ppb) at room temperature, until a stable baseline was reached. Gas mixtures were prepared by admixing certified gas standards to dry synthetic air with high-precision and calibrated mass flow controllers (Bronkorst, Netherlands). The standards (all from PanGas, Switzerland) were: benzene (15 ppm), ethanol (15 ppm), $H_2$ (47.9 ppm), acetone (18.1 ppm), NO (10.7 ppm), $NO_2$ (23 ppm), toluene (15 ppm), and xylene (10.3 ppm, all in synthetic air). Humidified air was generated by bubbling dry synthetic air through deionized water at 23 °C. The relative humidity was adjusted by dosing humidified air into the analyte-containing gas stream, as controlled with a SHT2x sensor (Sensirion, Switzerland). The total flow rate was 1 l/min. All transfer lines were made of Teflon and heated to 55 °C to mitigate water condensation and analyte adsorption.

The sensors were mounted onto the gas mixing setup directly after the conversion finished and were left for at least 2 h at 75 °C in a flow of 1 l/min at 50% RH until their baseline stabilized, prior to exposing them to the analytes. The response of the $Cu_3N$ sensor is defined as $S = (R_{\text{analyte}} - R_{\text{air}})/R_{\text{air}}$ for a reducing gas and as $S = (R_{\text{air}} - R_{\text{analyte}})/R_{\text{analyte}}$ for an oxidizing gas, where $R_{\text{analyte}}$ is the steady state resistance of the sensing film after exposure to the analyte and $R_{\text{air}}$ is the steady state resistance of the sensor in air. The response and recovery time are defined as the times required to reach 90 % of the response and the time to recover from 90 % of the resistance change, respectively. The selectivity is defined as the ratio between the response to $NO_2$ and the response to a confounder at equivalent concentrations.

## Data availability

The data that support the findings of this study can be requested from the corresponding author.

## Acknowledgment

This study was financially supported by the Swiss State Secretariat for Education, Research and Innovation (SERI) under contract number MB22.00041 (ERC-STG-21 "HEALTHSENSE"). The authors acknowledge Dr. Milivoj Plodinec from the Scientific Center for Optical and Electron Microscopy (ScopeM) of ETH Zurich for providing measuring time on their electron microscopes.

## Conflict of interest

The authors declare no conflict of interest.

## Author contributions

**Adrien Baut:** Conceptualization, Methodology, Investigation, Data Curation, Writing – Original Draft, Writing – Review & Editing, Visualization. **Michael Pereira Martins:** Methodology, Investigation, Data curation, Writing- Original Draft. **Andreas T. Güntner:** Conceptualization, Methodology, Writing – Review & Editing, Visualization, Supervision, Funding Acquisition

# Additional information

**Supplementary information** is available for this paper.

**Correspondence and requests for materials** should be addressed to A.T.G.

# Figures, Tables and captions

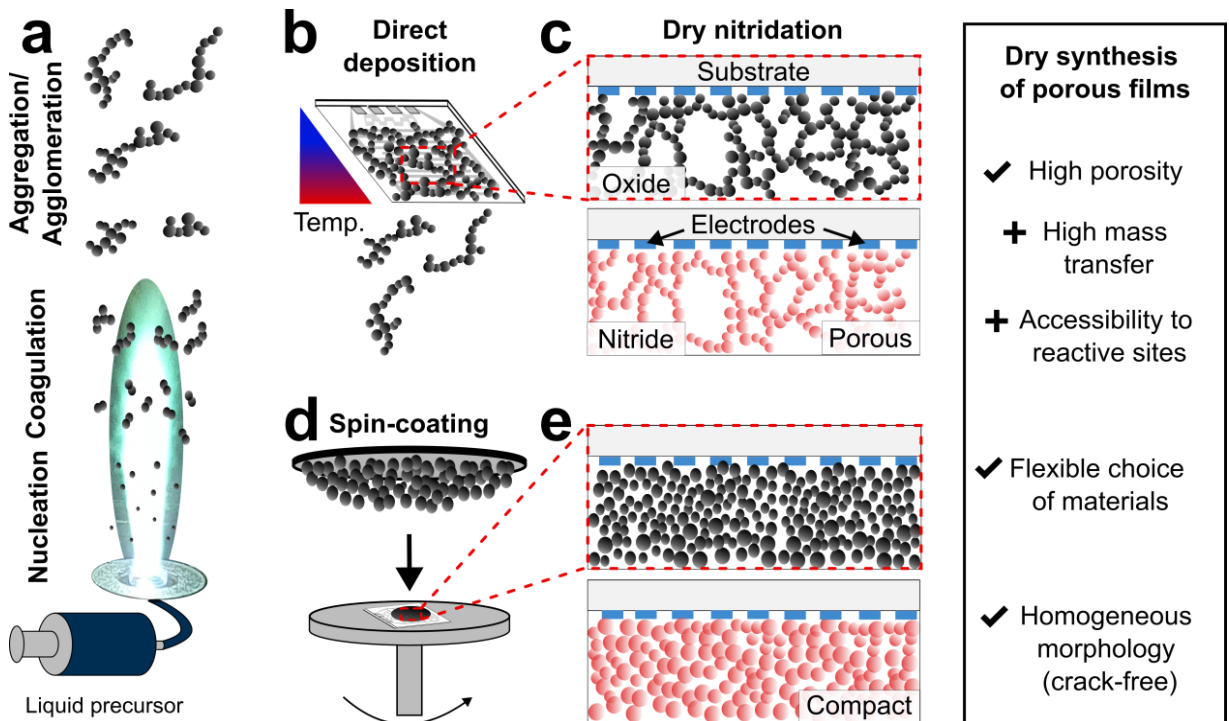

**Figure 1. Template-free and dry synthesis of porous MN architectures.** (a) Picture of a spray flame and illustrated formation of MOx nanoparticle agglomerates. Note that the shown flame combusted a Cu-containing precursor, but the process is versatile, as additionally shown for W, Ti, and Mo. (b) Direct deposition on substrates via thermophoresis and (c) their subsequent dry nitridation, yielding porous MN, as schematically illustrated. (d) For comparison, denser CuO films were prepared via spin-coating of similarly produced and filter-collected particles and (e) subjected to the same dry nitridation process than the porous films.

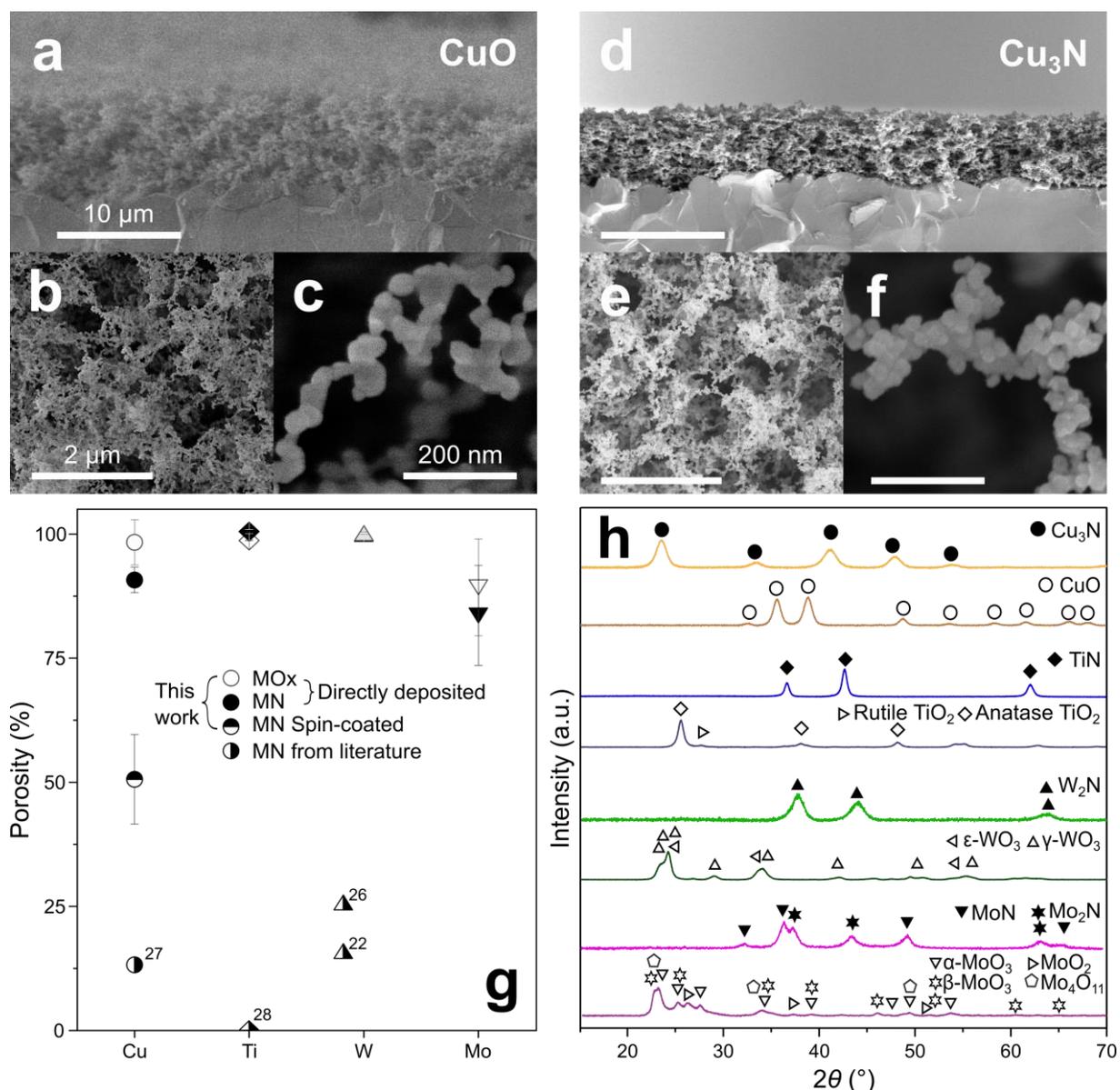

**Figure 2. Characterization of MN porous films and literature comparison.** (a) Cross-sectional and (b,c) top-view SEM images of highly porous CuO films. (d,e,f) The corresponding images of a $Cu_3N$ film. The scale bars in (d,e,f) match those in (a,b,c) respectively. (g) The porosity of directly deposited MOx (empty symbols) and MN (filled). For comparison, a spin-coated $Cu_3N$ (upper-half filled circles) and sputtered MN from literature (right-half filled symbols) are shown. (h) XRD patterns of MOx powders and the corresponding MN. The filled symbols are assigned to MN phases, while the empty symbols to MOx ones (see Methods for description of reference phases).

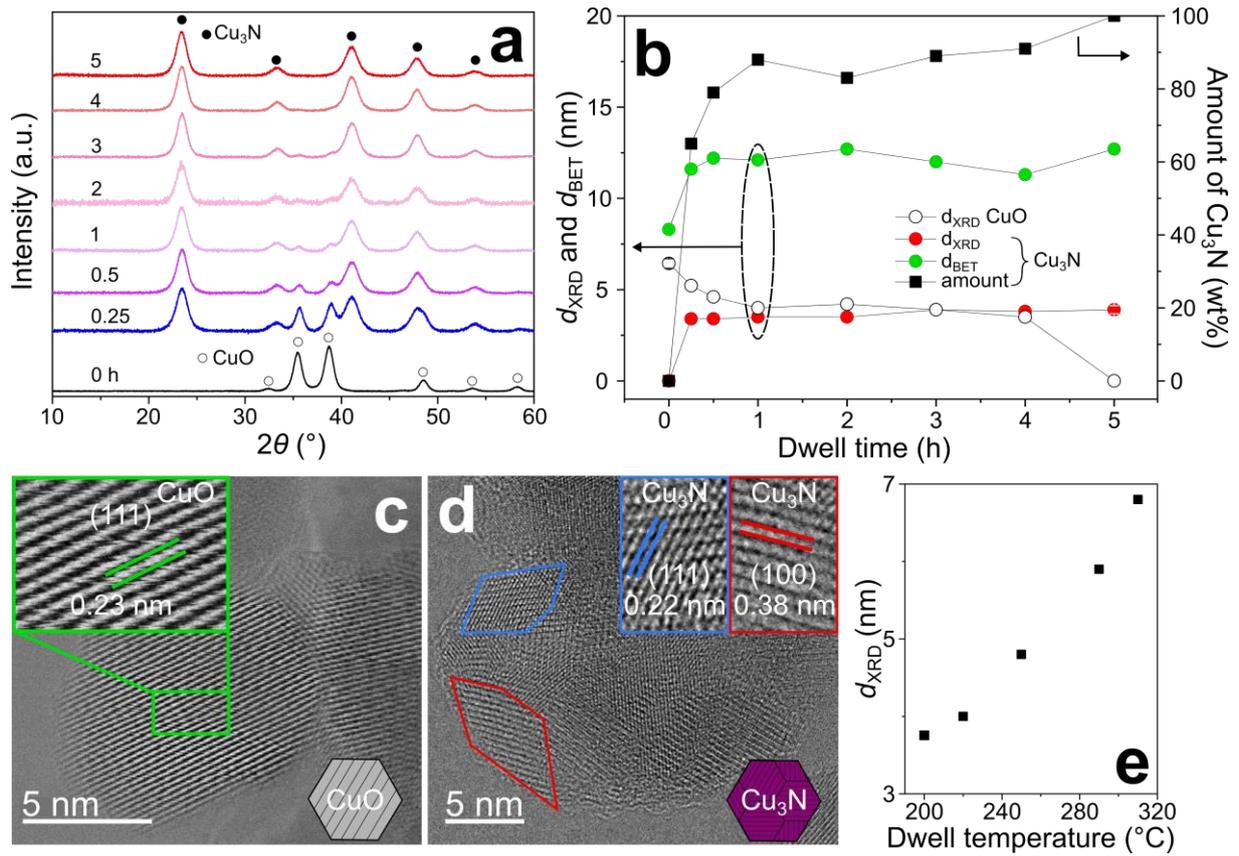

**Figure 3. Nanoparticle conversion of CuO into Cu$_3$N.** (a) XRD of CuO powder exposed to NH$_3$ at 200 °C for varying dwell times. (b) Corresponding crystal size of CuO (empty circles) and Cu$_3$N (red circles) with the BET-equivalent particle size (green circles) shown on the left and the amount (wt%) of Cu$_3$N on the right ordinate. For the crystal size of the CuO and the Cu$_3$N after 5 h, the symbols and error bars (hidden behind symbols) indicate average values and standard deviations from N = 3 identically prepared samples. TEM images of (c) CuO and (d) Cu$_3$N particles converted at 200 °C for 5 h. Insets provide higher resolution images with determined lattice spacing and corresponding lattice orientation for CuO (c) and Cu$_3$N (d), respectively. The illustrations represent the mono- and polycrystallinity of the CuO and the Cu$_3$N particles, respectively. (e) Crystal size of Cu$_3$N converted for 2 h at different dwell temperatures.

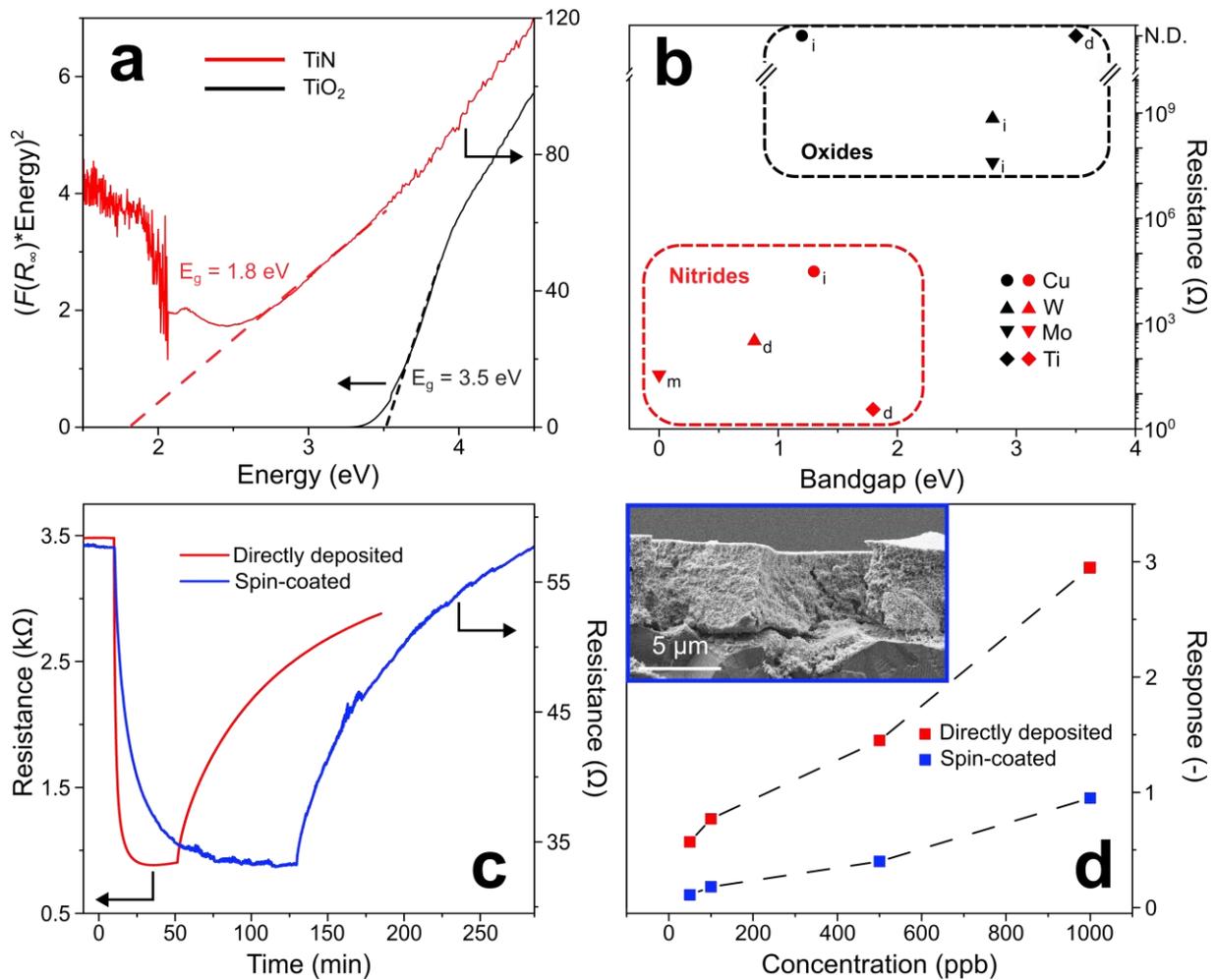

**Figure 4. Electronic and molecular sensing effect of dry nitridation.** (a) Tauc plot with the Kubelka-Munk function $F(R_\infty)$ for TiN (red) and $TiO_2$ (black). The dashed lines extrapolate the linear part of the curves for the determination of the bandgap. (b) Resistance of the directly deposited MOx (black) and MN films (red) obtained by dry conversion. Note that the resistances for the CuO and $TiO_2$ were too high (i.e. N.D.) for the instrument (> 1.2 GΩ) at room temperature. The subscripts refer to the type of bandgap: i for indirect, d for direct and m for metallic. (c) Resistance change over time of a directly deposited (red) and spin-coated $Cu_3N$ (blue) sensor at 75 °C during an exposure to 1000 ppb of $NO_2$ at 50% RH. (d) The corresponding response of the two sensors over a wide range of concentrations. The inset shows a cross-sectional SEM image of the spin-coated (blue) $Cu_3N$ sensor, while those of directly deposited were shown in Figure 2d-f.

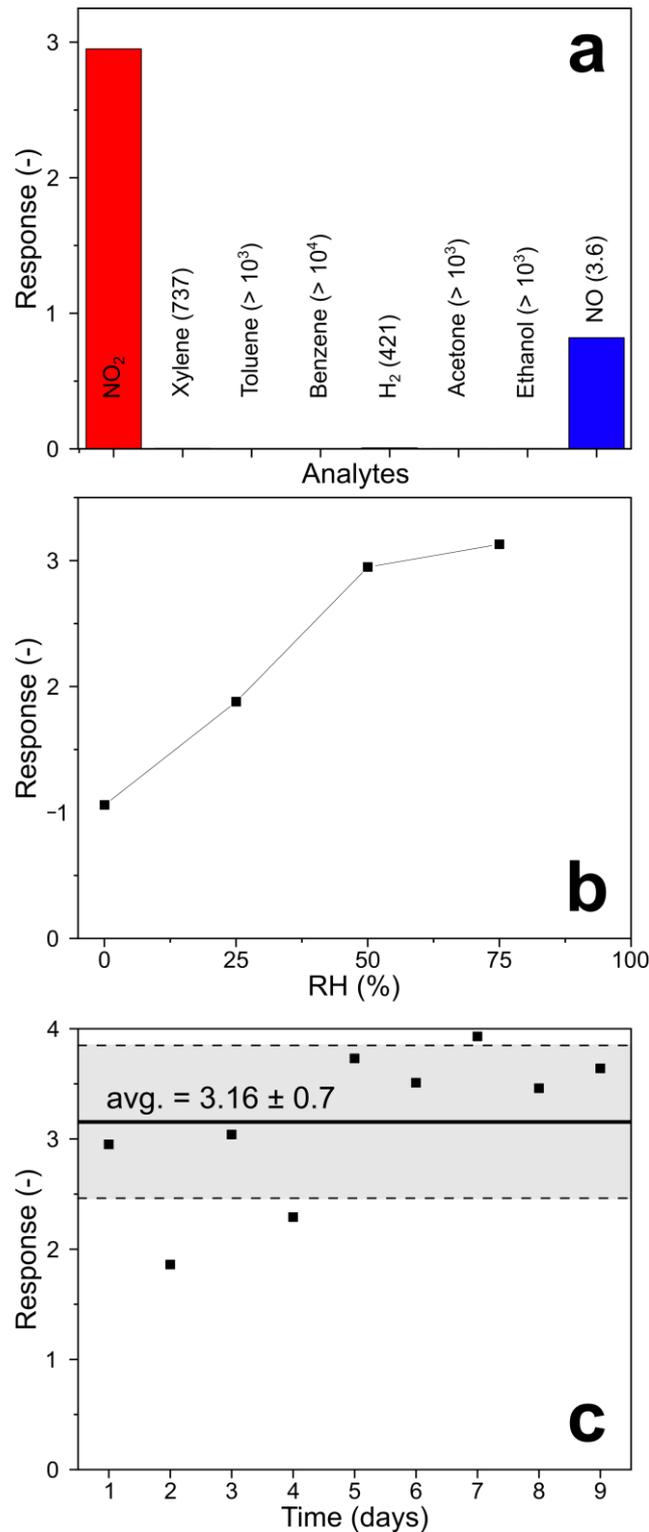

**Figure 5**: **Extended sensor characterization of porous Cu$_3$N.** (a) Response over a wide range of analytes (all tested at 1000 ppb) of a directly deposited Cu$_3$N sensor operated at 75 °C in air at 50% RH. The selectivity values are indicated in brackets. (b) Response under different RH and (c) over 9 days of daily exposure to 1000 ppb of NO$_2$. The solid line indicates the average response value (avg.) and the dashed lines ± standard deviation.

# Supplementary Information

## Porous metal nitride film synthesis without template

Adrien Baut[1], Michael Pereira Martins[1] and Andreas T. Güntner[1,*]

[1] Human-centered Sensing Laboratory, Department of Mechanical and Process Engineering, ETH Zurich, CH-8092 Zurich, Switzerland.

## Contents

# Supplementary Tables & Figures

**Table S1**: Crystal sizes of MOx and the respective MN. Note that the crystal sizes for $MoO_3$ are not included due to many overlapping peaks in the XRD pattern.

| Element | | CuO | $TiO_2$ | | $WO_3$ | | $MoO_3$ | |
|---|---|---|---|---|---|---|---|---|
| Phase | | | rutile | anatase | ε | γ | $Mo_2N$ | MoN |
| Crystal size (nm) | Oxide | 6.4 | 4.7 | 8.5 | 8.2 | 8.5 | - | |
| | Nitride | 3.9 | 12.3 | | 6.1 | | 9.6 | 8.0 |

**Table S2**: Precursor recipes and deposition parameters used for the synthesis of MOx powders and films.

| MOx | Molarity (M) | Precursor formulation | Pressure drop (bar) | Precursor flowrate (ml/min) | Deposition time (min) | Ref. |
|---|---|---|---|---|---|---|
| CuO | 0.25 | Deca Copper 8 (Borchers, Germany) in 2:1 (v/v) mixture of 2-ethylhexanoic acid (Sigma-Aldrich, Switzerland) and xylene (mixture of isomers, VWR Chemicals, Switzerland) | 1.6 | 4 | 9 | [53] |
| $TiO_2$ | 0.5 | Titanium tetraisopropoxide (TTIP, Aldrich, purity > 97%) in xylene (mixture of isomers, VWR Chemicals, Switzerland) | 2 | 5 | 4 | [84] |
| $WO_3$ | 0.2 | Ammonium metatungstate hydrate (Sigma-Aldrich, Switzerland) in 1:1 (v/v) mixture of ethanol (Sigma-Aldrich, Switzerland) and diethylene glycol monobutyl ether (DGME, Sigma-Aldrich, Switzerland) | 1.5 | 5 | 4 | [45] |
| $MoO_3$ | 0.2 | Ammonium molybdate tetrahydrate (Aldrich, ≥ 99%) in 1:1 (v/v) mixture of ethylene glycol (Fluka, ≥ 99.5%) and ethanol (VWR Chemicals, Switzerland) | 1.6 | 5 | 4 | [50] |

**Table S3**. Conversion parameters used for the synthesis of MN.

| MN | Mass of MOx powder (mg) | Conversion temperature (°C) | Heating rate (°C/min) | Dwell time (h) | $NH_3$ flow (ml/min) |
|---|---|---|---|---|---|
| $Cu_3N$ | 60 | 310 | 10 | 2 | 75 |
| TiN | 80 | 800 | 20 °C/min until 600 °C, then 3 °C/min | 7.5 | 75 |
| $W_2N$ | 60 | 600 | 10 | 1 | 75 |
| $MoN_x$ | 60 | 700 | 5 | 2 | 160 |

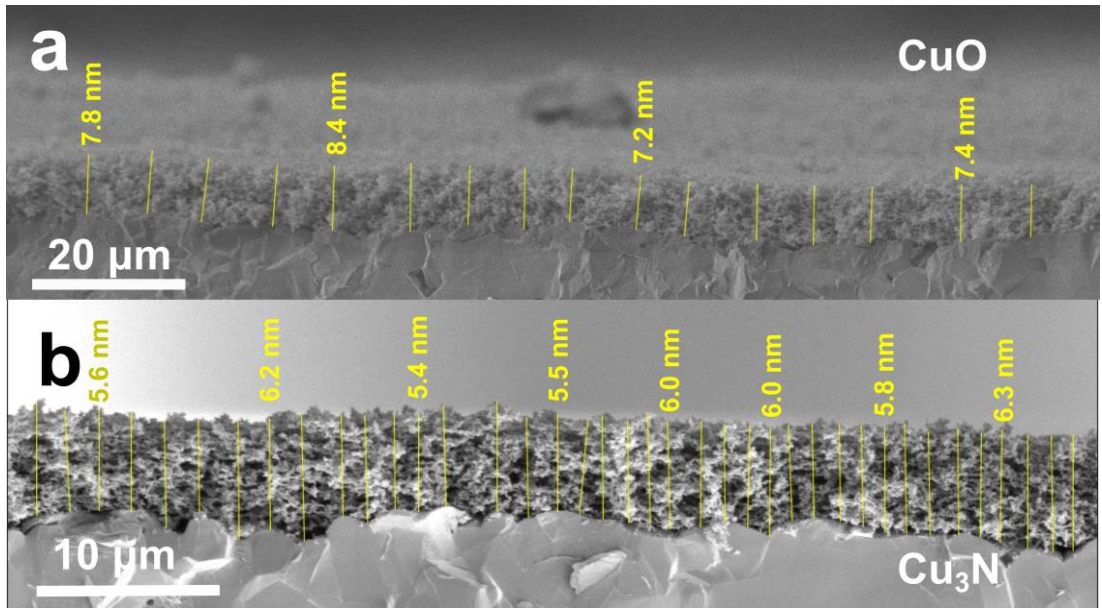

**Figure S1. Thickness of directly deposited CuO and Cu₃N films.** SEM images of the cross-section of a directly deposited (a) CuO and (b) dry converted Cu$_3$N on an Al$_2$O$_3$ substrate. The yellow lines (N = 16 for (a) and N = 38 for (b)) indicate examples (out of > 100) of measured film thicknesses.

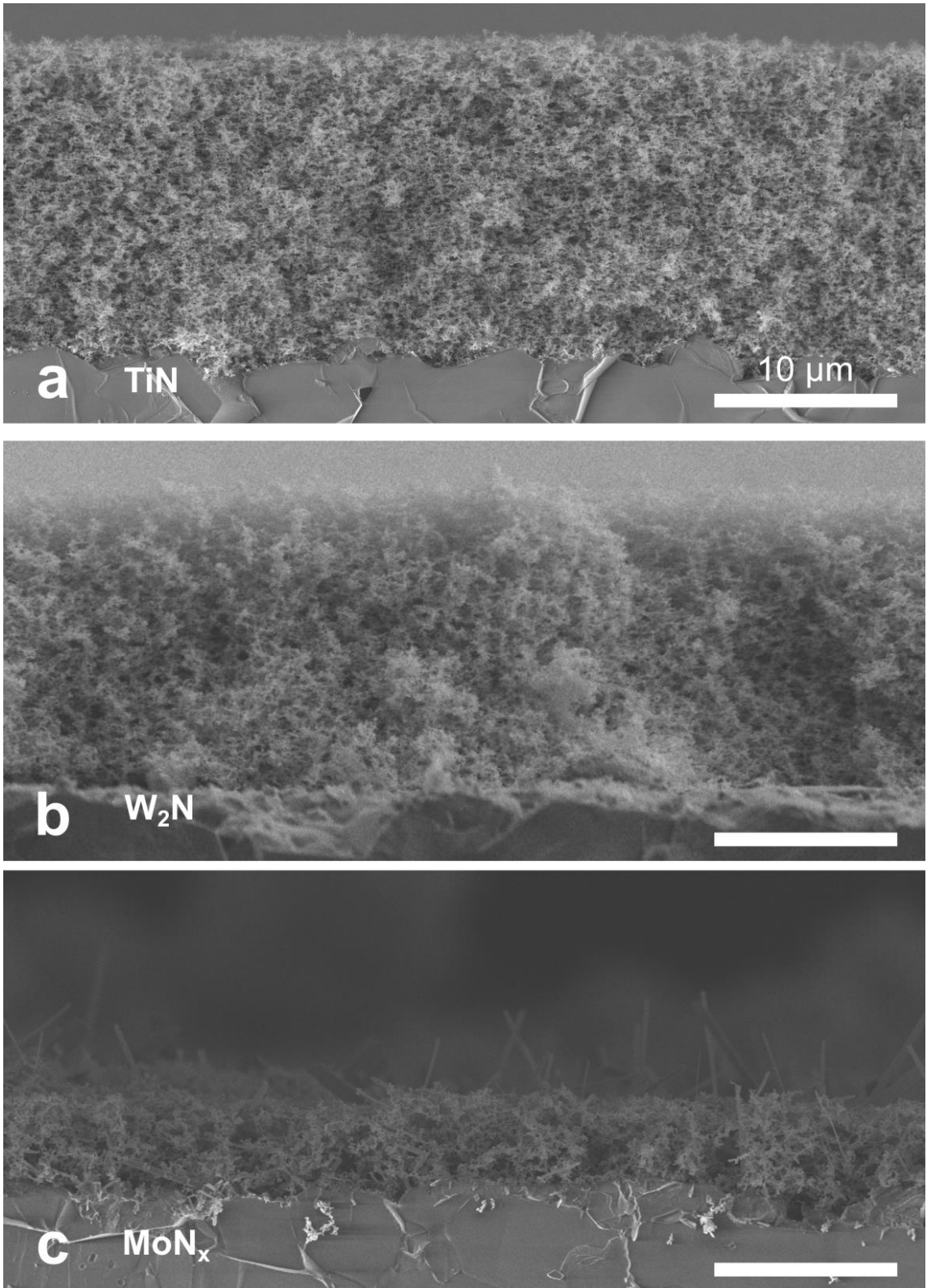

**Figure S2. Porous MN films obtained after dry nitridation.** SEM image of the cross-section of directly deposited and dry converted films of a) TiN, b) $W_2N$ and c) $MoN_x$. The scale bars represent 10 µm.

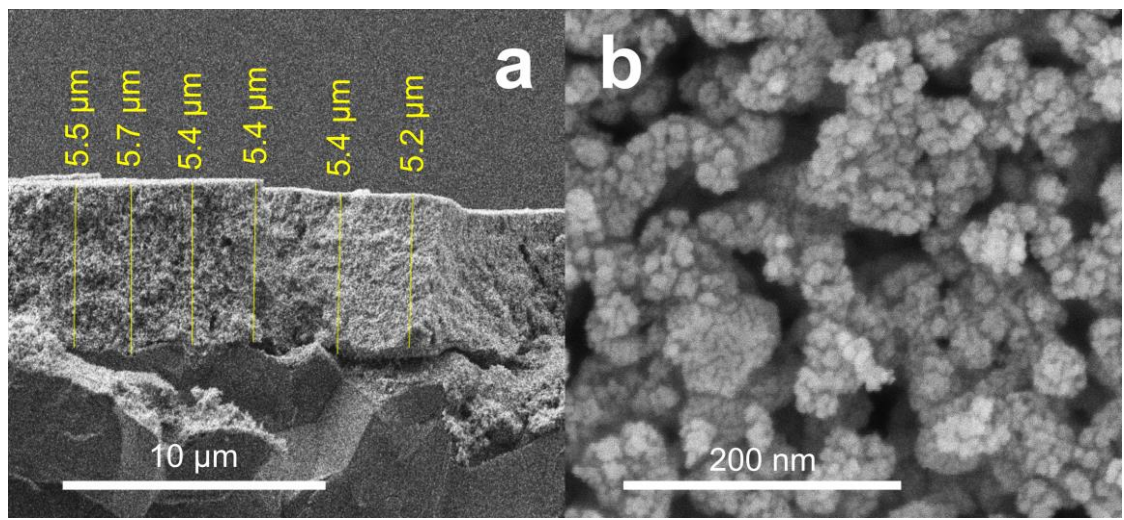

**Figure S3. MN films deposited by spin-coating.** SEM images of (a) the cross-section and (b) the top-view of a $Cu_3N$ film deposited onto an $Al_2O_3$ substrate by spin-coating and converted into $Cu_3N$. The yellow lines (N = 6) indicate examples (out of > 100) of measured film thicknesses.

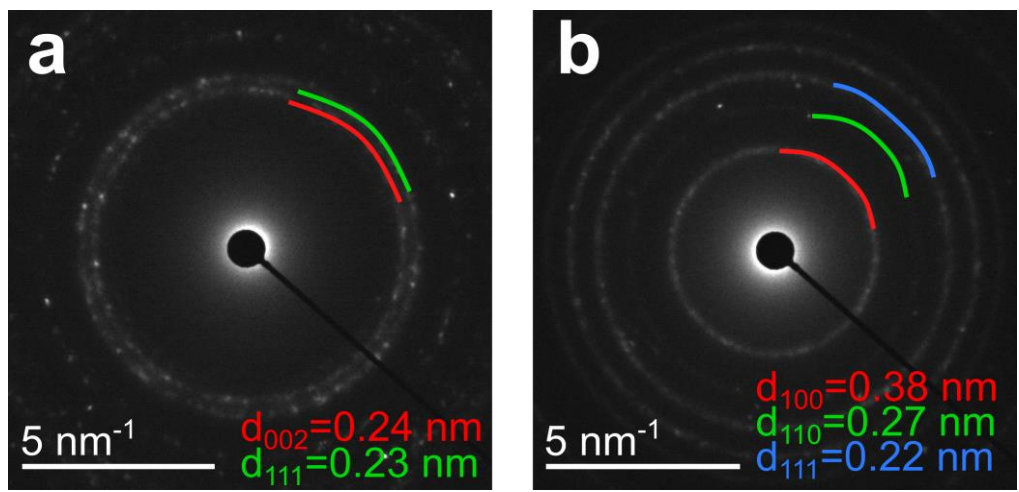

**Figure S4. SAED patterns of CuO and Cu$_3$N.** (a) SAED image of as-prepared CuO particles. The diffraction patterns are attributed to the (002) (red) and (111) (green) crystal orientations, respectively. (b) SAED pattern of the Cu$_3$N obtained by dry conversion for 5 h at 200 °C. The crystal planes identified are (100) (red), (110) (green) and (111) (blue).

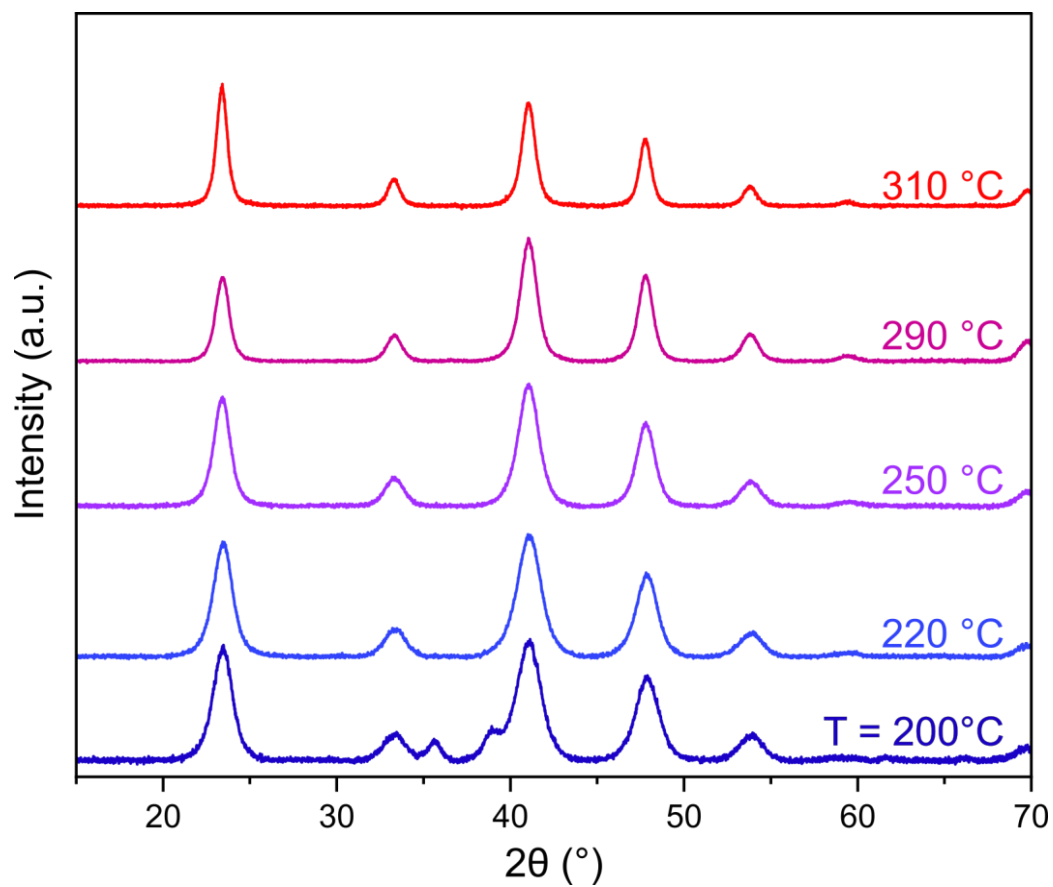

**Figure S5. XRD patterns of Cu₃N converted at different temperatures.** XRD patterns of Cu$_3$N powders after 2 h conversion at temperatures ranging from 200 °C – 310 °C.

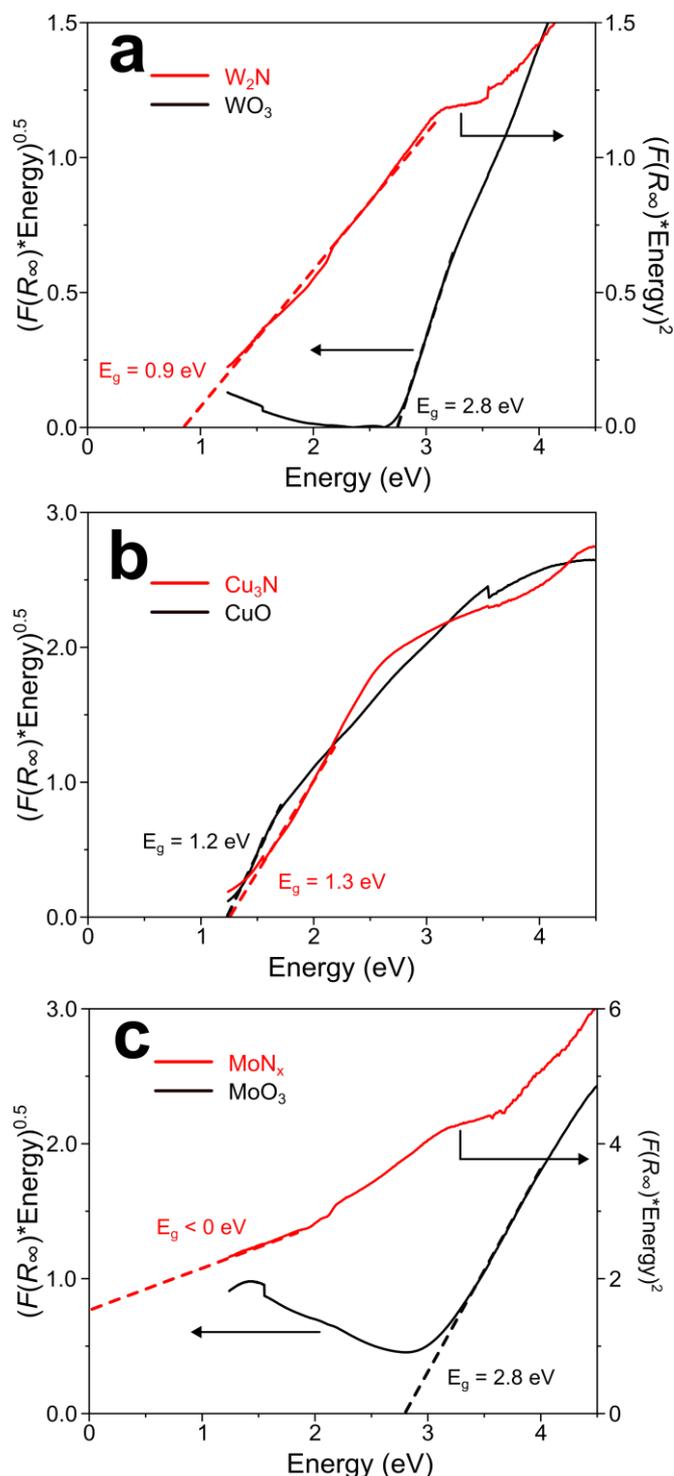

**Figure S6. Bandgap values of MN and their corresponding MOx.** Tauc plots using the Runge-Kutta function of (a) $W_2N$ and $WO_3$, (b) $Cu_3N$ and $CuO$, and (c) $MoN_x$ and $MoO_3$. The value $1/\gamma = 2$ in the function $(F(R_\infty) * \text{Energy})^{1/\gamma}$ indicates a direct bandgap, while $1/\gamma = 0.5$ an indirect transition The bandgap ($E_g$) is determined by extrapolating the linear part of the curves to an ordinate of 0. In the case of $MoN_x$, that value is negative, showing its metallic behaviour.

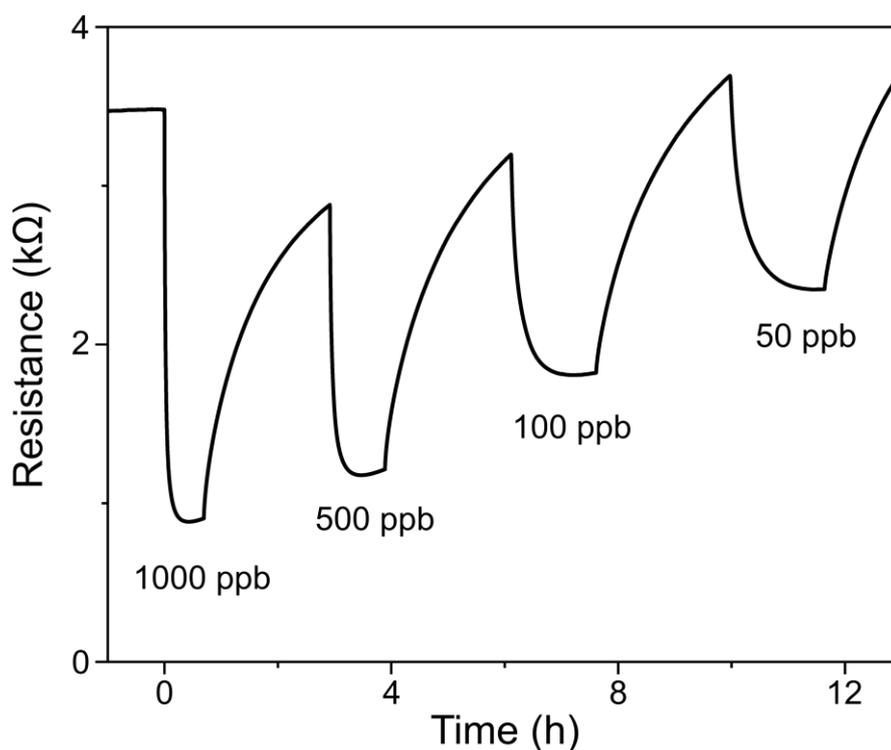

**Figure S7: Exposure of a porous Cu₃N sensor to different NO₂ concentrations.**

Resistance over time of a porous $Cu_3N$ sensor operated at 75 °c in air at 50% RH, while exposed at $NO_2$ concentrations ranging from 1000 – 50 ppb.